\documentclass[a4paper,11pt]{article}
\usepackage{pos}
\usepackage{float}
\usepackage{flexisym}
\usepackage{lmodern}

\title{Event characterization of dark bosons via exotic Higgs decays with final states of displaced dimuons in high luminosity era of the LHC}

\author*[a]{Tamer Elkafrawy}
\author[a]{Marcus Hohlmann}
\author[b]{Teruki Kamon}
\author[c]{Paul Padley}

\affiliation[a]{Florida Institute of Technology,\\
Melbourne, Florida USA}
\affiliation[b]{Texas A\&M University,\\
College Station, Texas USA}
\affiliation[c]{Rice University,\\
Houston, Texas USA}

\emailAdd{tamer.elkafrawy@cern.ch}
\emailAdd{telkafrawy@fit.edu}
\emailAdd{taelkafr@fnal.gov}

\abstract{We investigate the potential reach of a search for a long-lived/prompt dark vector boson $Z_D$, also called dark $Z$, and a prompt dark Higgs boson $h_D$ through exotic decays of the observed Higgs boson $h$ into either $Z_DZ_D$, $h_Dh_D$, or $ZZ_D$ with $Z$ being the hypercharge gauge boson. The $Z_D$ production through the Higgs portal is completed via one of two mechanisms, kinetic mixing of $Z_D$ with $Z$ and the mixing of $h_D$ with $h$. All production modes of $h$ are considered, while the branching fractions are calculated in Monte Carlo simulation using the {\textsc{MadGraph5}}\_aMC@NLO v2.7.2 framework. We focus on a final state of multiple dimuons, displaced up to \mbox{7500 mm}, where the muons can be reconstructed without vertex constraint using data from ATLAS and CMS detectors to be collected in Run~3. Integrated luminosities of 137, 300, and 3000 fb$^{-1}$ for Run~2, Run~3, and high luminosity run (HL-LHC), respectively, are used for estimating the expected search sensitivity of the Large Hadron Collider to each of the decay modes.}

\begin{document}
\maketitle

\section{Introduction}
The observed Higgs boson $h$ plays a significant role in the Standard Model (SM) and is believed to impact a wide range of new physics beyond the SM (BSM). It is assumed that $h$ is responsible for breaking the electroweak symmetry and that there is an additional $U(1)_D$ dark gauge symmetry allowing $h$ to decay to new particles such as dark Higgs and dark vector bosons, usually referred to as $h_D$ and $Z_D$, respectively. The only possible interaction of $Z_D$ with the SM sector is through its kinetic mixing (KM) with $Z$ boson, while if the Higgs mixing (HM) exists, $h_D$ will have a renormalizable coupling to $h$. The high integrated luminosities ${\cal L}$'s achieved by the Large Hadron Collider (LHC) offer a promising opportunity into the search for hidden sectors through these two portals.

\begin{figure}[H]
\includegraphics[width=1.00\textwidth]{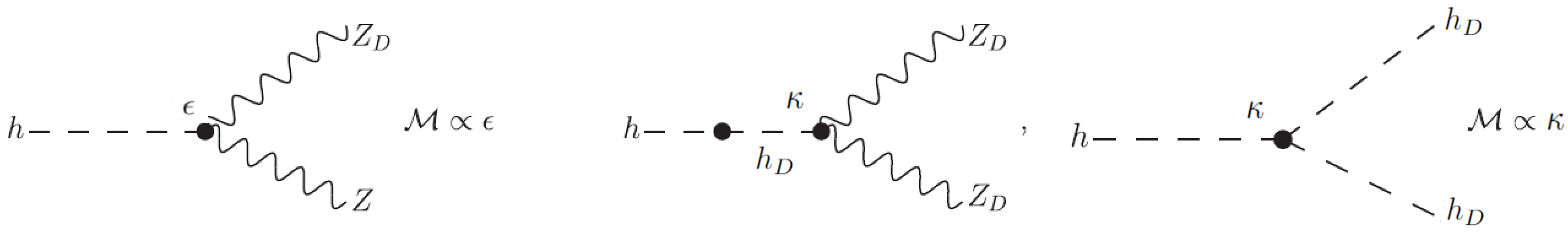}
\caption{\label{fig1}Feynman diagrams for the dominant exotic Higgs decays via KM (left) and HM (middle and right) with $\cal M$ being the Higgs decay matrix element and $\epsilon$ and $\kappa$ being the kinetic and Higgs mixing parameters, respectively \cite{Curtin2014}. The vertex $hh_DZ_D$ exists but is highly suppressed by each of the two types of mixing.}
\end{figure}

Muons are efficiently identified in ATLAS and CMS detectors, and hence we assume the muons can be reconstructed without vertex constraint with kinematic acceptances and efficiencies of 100\%. This simplifies our two-dimensional scans over the relevant free parameters of the dark sector using all the current NLO simulated observables and enables us to characterize exotic decays of $h$ by coming up with the most stringent constraints on such free parameters, which are the KM parameter $\epsilon$, HM parameter $\kappa$, and the acquired masses by $Z_D$ and $h_D$, denoted by $m_{Z_D}$ and $m_{h_D}$, respectively. Feynman diagrams of the three exotic decay modes are given in Fig.~\ref{fig1}. In this context, a mass of $m_h$ = 125.09~GeV is considered \cite{Cepeda2019}, which is assumed to be produced at the LHC through the production modes of gluon-gluon fusion $(gg$F), vector-boson fusion (VBF), $VH$ (i.e., $W^+h$, $W^-h$, $\ell^+{\nu}h$, $\ell^-\overline{\nu}h$, $Zh$, $\ell\overline{\ell}h$, ${\nu}\overline{\nu}h$), $t\overline{t}h$, $th$, and $\overline{t}h$ with a production cross section of 55.88 and 63.06 pb for 13 and 14 TeV, respectively, which are calculated to either next-to-leading order with QCD corrections included (NLO QCD), next-to-next-to-leading order with QCD corrections included (N$^{2}$LO QCD), or next-to-next-to-next-to-leading order with QCD corrections included (N$^{3}$LO QCD), combined or not combined with next-to-leading order with electroweak corrections included (NLO EW), depending on the production mode \cite{Cepeda2019}. The simulated samples for this work are generated by applying Monte Carlo (MC) simulation using the framework of {\textsc{MadGraph5}}\_aMC@NLO v2.7.2 with Hidden Abelian Higgs Model (HAHM) \cite{Curtin2014}.

\section{Sensitivity of the LHC to the searches for long-lived dark vector bosons}

\begin{figure}[H]
\includegraphics[width=0.33\textwidth]{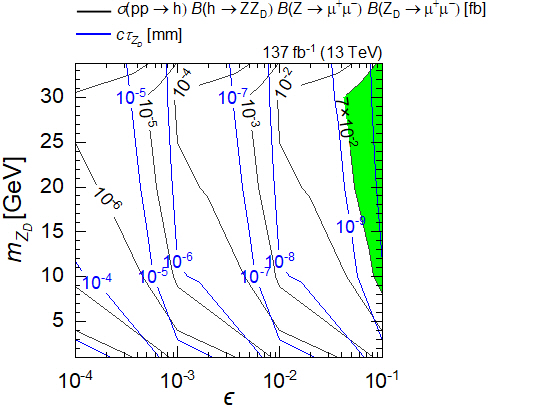}
\includegraphics[width=0.33\textwidth]{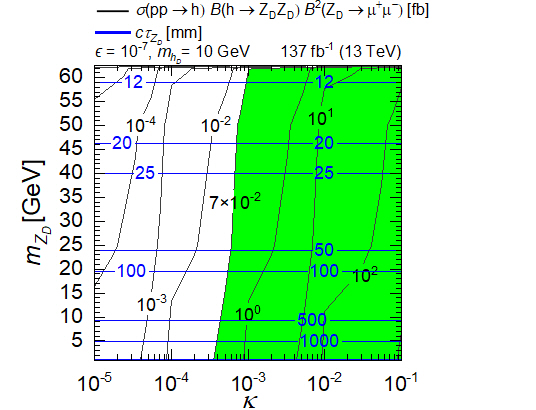}
\includegraphics[width=0.33\textwidth]{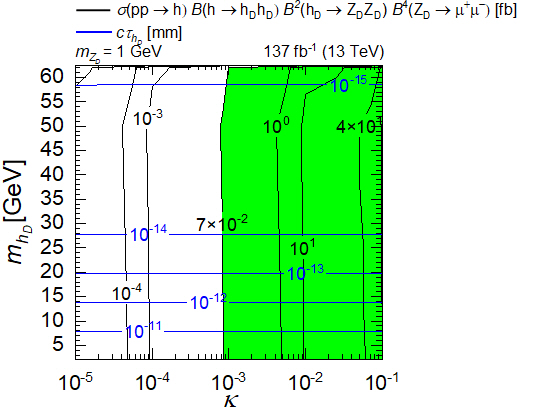}
\includegraphics[width=0.33\textwidth]{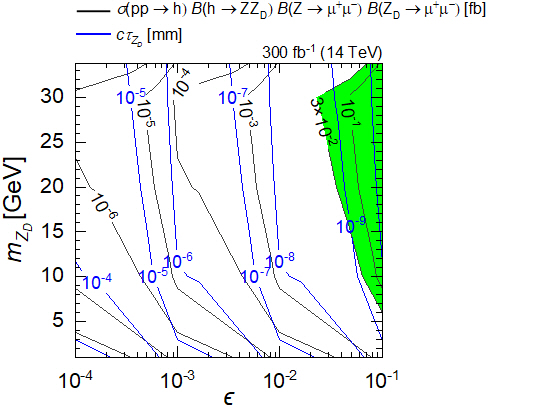}
\includegraphics[width=0.33\textwidth]{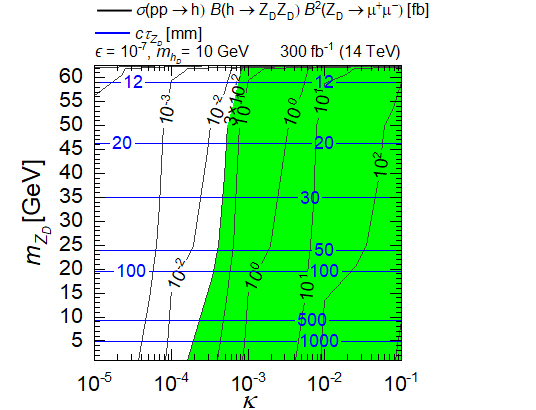}
\includegraphics[width=0.33\textwidth]{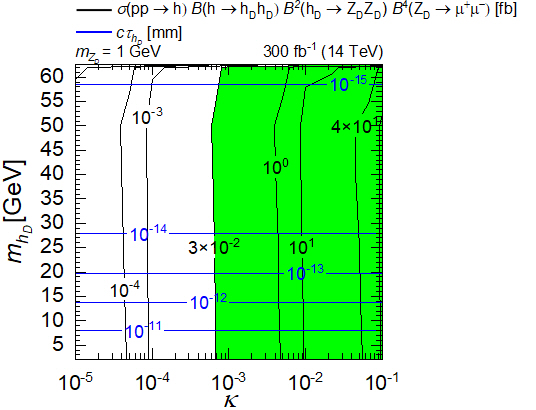}
\includegraphics[width=0.33\textwidth]{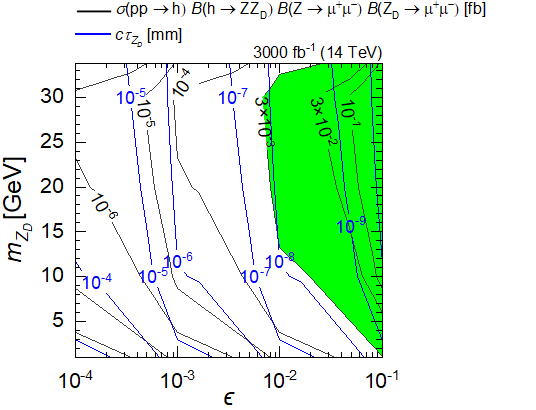}
\includegraphics[width=0.33\textwidth]{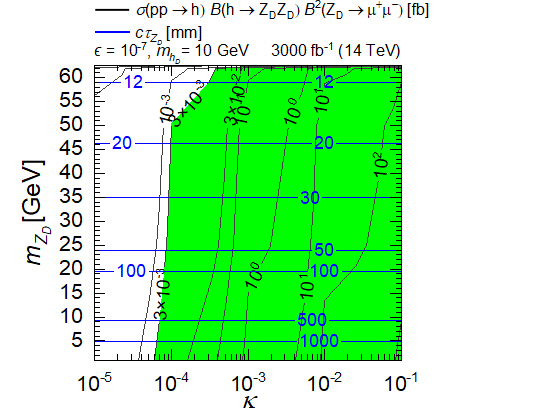}
\includegraphics[width=0.33\textwidth]{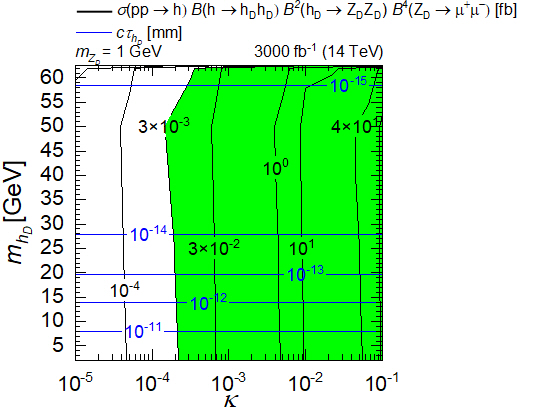}
\caption{\label{fig2}MC simulations showing the contour lines of $\sigma_{total}$ (black) and $c\tau_{Z_D}$ (or $c\tau_{h_D}$) (blue) for $h\rightarrow{ZZ_D}\rightarrow{2\mu^{+}2\mu^{-}}$ (left column), $h\rightarrow{Z_DZ_D}\rightarrow{2\mu^{+}2\mu^{-}}$ (middle column), and $h\rightarrow{h_Dh_D}\rightarrow{4Z_D}\rightarrow{4\mu^{+}4\mu^{-}}$ (right column) in a scan over the $\epsilon$-$m_{Z_D}$, the $\kappa$-$m_{Z_D}$, and the $\kappa$-$m_{h_D}$ planes, respectively, for Run~2 (top row), Run~3 (middle row), and HL-LHC (bottom row) with sensitivity regions shaded in green.}
\end{figure}

The observed Higgs boson $h$ is considered to be produced through all possible production modes with production cross sections of 55.88 and 63.06~pb for 13 and 14~TeV, respectively. The values of 0.073, $0.0\overline{3}$, and $0.00\overline{3}$~fb are taken as the smallest $\sigma_{total}$ to which the LHC is sensitive based on 10 events at least to be measured in Run~2, Run~3, and HL-LHC with a full ${\cal L}$'s of 137, 300, and 3000~fb$^{-1}$, respectively, and accordingly the contour lines of $\sigma_{total}$ are shown in all panels of Fig.~\ref{fig2} where $\sigma_{total}$ equals to $\sigma(pp\rightarrow{h})~{\cal B}(h\rightarrow{ZZ_D})~{\cal B}(Z\rightarrow{\mu^+\mu^-})~{\cal B}(Z_D\rightarrow{\mu^+\mu^-})$, $\sigma(pp\rightarrow{h})~{\cal B}(h\rightarrow{Z_DZ_D})~{\cal B}^2(Z_D\rightarrow{\mu^+\mu^-})$, and $\sigma(pp\rightarrow{h})~{\cal B}(h\rightarrow{h_Dh_D})~{\cal B}^2(h_D\rightarrow{Z_DZ_D})~{\cal B}^4(Z_D\rightarrow{\mu^+\mu^-})$ in the case of $h\rightarrow{ZZ_D}\rightarrow{2\mu^{+}2\mu^{-}}$, $h\rightarrow{Z_DZ_D}\rightarrow{2\mu^{+}2\mu^{-}}$, and $h\rightarrow{h_Dh_D}\rightarrow{4Z_D}\rightarrow{4\mu^{+}4\mu^{-}}$, respectively. The contour lines of $c\tau_{Z_D}$ and $c\tau_{h_D}$ are superimposed on those of $\sigma_{total}$ with any consecutive lines separated by one order of magnitude. The chosen value of $\epsilon=10^{-7}$ and the $m_{Z_D}$ region of $1-m_h/2\approx62.5$~GeV correspond to $c\tau_{Z_D}$ in the range of $10-2000$~mm for $h\rightarrow{Z_DZ_D}\rightarrow{2\mu^{+}2\mu^{-}}$ as seen in the middle column of Fig.~\ref{fig2}.

\section{Lifetime of dark bosons and impact on $\Gamma_W$ by the hidden sector via $h~\rightarrow~{Z_DZ_D}$}

\begin{figure}[H]
\includegraphics[width=0.474\textwidth]{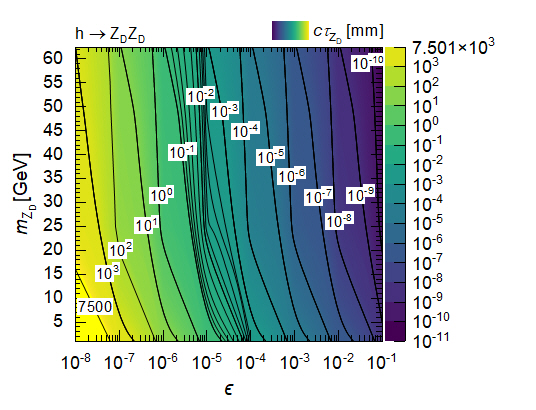}
\includegraphics[width=0.474\textwidth]{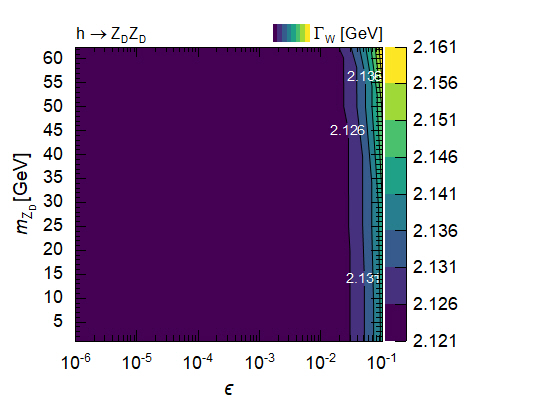}
\caption{\label{fig3} MC simulation of $c\tau_{Z_D}$ (left) and $\Gamma_W$ (right) as scanned over the $\epsilon$-$m_{Z_D}$ plane for $h\rightarrow{Z_DZ_D}$.}
\end{figure}

The decay length $c\tau_{Z_D}$ via $h\rightarrow~{ZZ_D}$ and $h\rightarrow~{Z_DZ_D}$ is fully described by the scan over the $\epsilon$-$m_{Z_D}$ plane, which is given by the left column of Fig.~\ref{fig2} and the left panel of Fig.~\ref{fig3}, respectively, where $c\tau_{Z_D}$ is seen to be inversely proportional to $\epsilon^2$ and $m_{Z_D}$. The inverse proportionality between $c\tau_{Z_D}$ and $m_{Z_D}$ is noticed by the middle column of Fig.~\ref{fig2} where a smaller $m_{Z_D}$ decays to as few as eight particles, $d$, $u$, $s$ quarks, $e^-$, $\mu^-$, $\nu_e$, $\nu_\mu$, and $\nu_\tau$ and in turn has a narrower decay width and longer $c\tau_{Z_D}$, while a heavier $Z_D$ can produce up to three more particles, $c$, $b$, and $\tau$, leading to a wider decay width and shorter $c\tau_{Z_D}$. On the other hand, the decay length $c\tau_{h_D}$ via $h\rightarrow~{h_Dh_D}$ is fully described by the scan over the $\kappa$-$m_{h_D}$ plane, which is given by the right column of Fig.~\ref{fig2}. The decay width of $W$ boson $\Gamma_W$ is found to change by $\sim2\%$ in the scan over the $\epsilon$-$m_{Z_D}$ plane where it is maximal for the highest values of $m_{Z_D}$ and $\epsilon$, and vice versa as seen in the right panel of Fig.~\ref{fig3}.

\section{Conclusion}

The LHC is found to be more sensitive to $h\rightarrow~{Z_DZ_D}\rightarrow{2\mu^{+}2\mu^{-}}$ and $h\rightarrow{h_Dh_D}\rightarrow{4Z_D}\rightarrow{4\mu^{+}4\mu^{-}}$, irrespective of the mass acquired by $Z_D$ and $h_D$, than $h\rightarrow~{ZZ_D}\rightarrow{2\mu^{+}2\mu^{-}}$. New constraints on KM and HM parameters are obtained for the first two decay modes in Run~2, Run~3, and HL-LHC as down to ($\kappa=3.5\times10^{-4}$, $1.5\times10^{-4}$, and $6.0\times10^{-5}$) and ($\kappa=8.5\times10^{-4}$, $6.5\times10^{-4}$, and $2.0\times10^{-4}$), respectively. For $h\rightarrow~{ZZ_D}\rightarrow{2\mu^{+}2\mu^{-}}$, the constraints are down to $\epsilon=4.0\times10^{-2}$ for the $m_{Z_D}$ range of $8.5-33.8$~GeV in Run~2, $\epsilon=2.0\times10^{-2}$ for the $m_{Z_D}$ range of $6.0-33.8$~GeV in Run~3, and $\epsilon=7.0\times10^{-3}$ for the $m_{Z_D}$ range of $1.0-33.8$~GeV in HL-LHC. The LHC is found to be sensitive to the production of prompt or long-lived $Z_D$'s ($10^{-10}-7500$~mm depending on $\epsilon$ and $m_{Z_D}$) via $h\rightarrow~{Z_DZ_D}\rightarrow{2\mu^{+}2\mu^{-}}$, while it is sensitive only to the production of prompt $Z_D$'s ($10^{-10}-10^{-8}$~mm) and prompt $h_D$'s ($10^{-15}-10^{-10}$~mm) via $h\rightarrow{ZZ_D}\rightarrow{2\mu^{+}2\mu^{-}}$ and $h\rightarrow{h_Dh_D}\rightarrow{4Z_D}\rightarrow{4\mu^{+}4\mu^{-}}$, respectively.

\begin{acknowledgments}

All institutions wish to thank the DOE's Office of Science (HEP) for its support of this work through the grants, DE-SC0013794, DE-SC0010103, and DE-SC0010813.
\end{acknowledgments}


\begin{thebibliography}{99}

\bibitem{Cepeda2019}
M. Cepeda, S. Gori, P. Ilten, M. Kado, and F. Riva, {Report from Working Group 2: Higgs physics at the HL-LHC and HE-LHC}, \href{https://doi.org/10.23731/CYRM-2019-007.221} {10.23731/CYRM-2019-007.221} \href{https://arxiv.org/abs/1902.00134} {[arXiv:1902.00134]}.

\bibitem{Curtin2014}
D. Curtin \emph{et al.}, {Exotic decays of the 125 GeV Higgs boson}, \href{https://doi.org/10.1103/PhysRevD.90.075004} {{Phys. Rev. D} {\bf90}, 075004 (2014)} \href{https://arxiv.org/abs/1312.4992} {[arXiv:1312.4992]}.

\end{thebibliography}
\end{document}